\documentclass[sigconf]{acmart}

\AtBeginDocument{%
  }

\setcopyright{acmlicensed}
\copyrightyear{2026}
\acmYear{2026}
\acmDOI{XXXXXXX.XXXXXXX}
\acmISBN{978-1-4503-XXXX-X/2018/06}

\usepackage{listings}
\usepackage{xcolor}
\usepackage{indentfirst}
\usepackage{enumitem}
\usepackage{tabularx}
\usepackage{pgfplots}
\usepackage{pgfplotstable}
\pgfplotsset{compat=1.18}
\usepackage{tikz}
\usetikzlibrary{decorations.pathreplacing,calc}
\usepackage{multirow}
\usepackage{makecell}
\usepackage{colortbl}

\begin{document}

\title{SmartEval: A Benchmark for Evaluating LLM-Generated Smart Contracts from Natural Language Specifications}

\author{Abhinav Goel}
\affiliation{%
  \institution{Columbia University}
  \city{New York}
  \state{NY}
  \country{USA}}
\email{ag5252@columbia.edu}

\author{Chaitya Shah}
\affiliation{%
  \institution{Columbia University}
  \city{New York}
  \state{NY}
  \country{USA}}
\email{cs4621@columbia.edu}

\author{Agostino Capponi}
\affiliation{%
  \institution{Columbia University}
  \city{New York}
  \state{NY}
  \country{USA}}
\email{ac3827@columbia.edu}

\author{Alfio Gliozzo}
\affiliation{%
  \institution{IBM T.J. Watson Research Center}
  \city{Yorktown Heights}
  \state{NY}
  \country{USA}}
\email{gliozzo@us.ibm.com}

\begin{abstract}
We introduce SmartEval, a benchmark for systematically evaluating the quality of Solidity smart contracts generated by large language models (LLMs) from natural language specifications. SmartEval provides a corpus of 9,000 generated contracts paired with expert-written ground-truth implementations drawn from the FSM-SCG dataset, a five-dimensional evaluation rubric covering functional completeness, variable fidelity, state-machine correctness, business-logic fidelity, and code quality, and a reproducible generation-and-evaluation pipeline.

To validate the benchmark's reliability, we conduct three independent empirical studies: a five-condition ablation study (N=300 per condition) isolating the contribution of each pipeline component, a human expert evaluation by three Columbia University PhD researchers confirming automated scores align with expert judgment to within 0.34 points, and external security analysis via the Slither static analyzer confirming 79.4\% agreement between the LLM auditor and a non-LLM rule-based tool. Systematic analysis of 9,000 generated contracts reveals characteristic failure modes (logic omissions at 35.3\%, state transition errors at 23.4\%, and complexity-driven degradation) and quantifies a +8.29 composite-score advantage of generated contracts over ground-truth implementations, attributable to LLMs' literal specification-following behavior. SmartEval establishes a reproducible, validated foundation for empirical research on LLM smart contract synthesis quality, with all data, evaluation code, and generated contracts publicly released.
\end{abstract}



\maketitle
\vspace{-10pt}

\section{Introduction}

Smart contracts, self-executing programs deployed on blockchain networks that automatically enforce agreement terms without intermediaries, have become critical infrastructure for decentralized finance (DeFi), non-fungible tokens (NFTs), and decentralized autonomous organizations (DAOs)~\cite{buterin2014next, wood2014ethereum, capponi2023defi}. However, developing secure and correct smart contracts remains a significant challenge due to the complexity of languages like Solidity, the immutability of deployed contracts, and the high-stakes nature of financial applications~\cite{atzei2017survey, chen2020survey}.

The gap between domain experts who understand contract requirements and developers who can implement them in Solidity creates bottlenecks in the development process. Traditional approaches require extensive back-and-forth between legal or business stakeholders and blockchain developers, often leading to specification misinterpretation and implementation errors. These errors have resulted in significant financial losses, with notable examples including the DAO hack (\$60M) and the Parity wallet freeze (\$300M)~\cite{luu2016making, mehar2017dao}. Recent work by Anthropic's red team~\cite{anthropic2025scone} demonstrated that frontier AI agents can now autonomously exploit real-world smart contract vulnerabilities, collectively extracting \$4.6M in simulated stolen funds from contracts exploited after model knowledge cutoffs, with exploit revenue doubling every 1.3 months.

Recent advances in Large Language Models (LLMs) have demonstrated remarkable capabilities in code generation~\cite{chen2021evaluating, austin2021program, roziere2024code}. Models such as GPT-4~\cite{openai2024gpt4} and Code Llama~\cite{roziere2024code} can generate syntactically correct code from natural language descriptions. However, generating production-ready smart contracts requires more than syntactic correctness; it demands semantic fidelity to specifications, proper state machine implementation, security best practices, and correct economic logic. Emerging techniques such as SmartInv~\cite{wang2024smartinv} and SmartSys~\cite{wang2024smartsys} leverage multimodal and foundation models to uncover ``machine un-auditable'' smart contract bugs that evade traditional static analysis. These results highlight both the promise and the urgency of developing systematic, validated quality evaluation frameworks for AI-generated smart contracts.

In this paper, we introduce \textbf{SmartEval}, a benchmark and evaluation framework for studying how well LLMs translate natural language contract specifications into correct, secure, and deployable Solidity code. SmartEval applies to a broad class of contract types including NDA agreements, employment contracts, rental agreements, service contracts, token standards, and governance systems, any domain where natural language must become verifiable on-chain logic. The benchmark provides three core assets: (i) a corpus of 9,000 LLM-generated contracts paired with 9,000 expert ground-truth implementations, (ii) a validated five-dimensional evaluation rubric with deterministic score computation, and (iii) a reproducible generation-and-evaluation pipeline with full provenance metadata. A complete worked example for a token staking contract is provided in Appendix~\ref{app:config} and Table~\ref{tab:example_evaluation}.

Our contributions include:

\begin{enumerate}
    \item \textbf{SmartEval Benchmark and Dataset}: A corpus of 9,000 LLM-generated Solidity contracts paired with expert ground-truth implementations, spanning six contract categories (token standards, governance, staking, escrow, NDA/legal, and general service contracts), with per-contract quality scores, security reports, and compilation results, all publicly released for reproducible evaluation.

    \item \textbf{Five-Dimensional Quality Evaluation Rubric}: The first rubric to treat FSM state machine correctness as a weighted first-class metric in LLM code generation evaluation, measuring Functional Completeness (25\%), Variable Fidelity (15\%), State Machine Correctness (15\%), Business Logic Fidelity (35\%), and Code Quality (10\%), with deterministic composite recomputation eliminating drift from model-generated aggregates.

    \item \textbf{Systematic Failure Mode Analysis}: Characterization of LLM error patterns across 2,398 lower-performing contracts, revealing that logic omissions (35.3\%) and state transition errors (23.4\%) are the dominant failure modes, with performance degrading sharply on high-complexity specifications (8+ functions, 5+ states: avg.\ score 71.8 vs.\ 87.2 for low-complexity).

    \item \textbf{Three-Way Benchmark Validation}: Independent confirmation of rubric reliability via (a) human expert scoring by three Columbia University PhD researchers (within 0.34 points of automated evaluator), (b) external Slither static analysis (79.4\% LLM--tool vulnerability category agreement), and (c) a five-condition ablation study with N=300 per condition and cross-seed Cohen's $d < 0.2$.

    \item \textbf{Generation Pipeline with Severity-Gated Refinement}: A reproducible multi-agent pipeline (parser, generator, auditor, refiner, evaluator) that produces all benchmark artifacts, with a control-flow security gate whose removal increases output standard deviation by 111\% and reduces compilation by 5.2 percentage points.

    \item \textbf{Behavioral Analysis of LLM vs.\ Expert Developers}: Quantitative characterization of the +8.29 composite gap between LLM-generated and expert-written contracts, attributing it to LLMs' literal specification-following behavior versus expert developers' architectural judgment, gas-efficiency trade-offs, and pattern reuse.
\end{enumerate}

\section{Background}

Smart contracts are programs stored on blockchain networks that execute automatically when predetermined conditions are met~\cite{nakamoto2008bitcoin, buterin2014next}. Ethereum introduced Solidity as its primary programming language~\cite{wood2014ethereum, solidity2024}. Key features include state variables, functions with visibility and mutability modifiers, access-control modifiers, event logging, and contract inheritance. The immutability of deployed contracts makes correctness paramount: bugs cannot be patched post-deployment without complex migration procedures~\cite{zheng2020overview}.

Common vulnerability classes include reentrancy attacks, integer overflow, access control flaws, and timestamp manipulation. Static analysis tools like Slither~\cite{slither2019} and Mythril~\cite{mythril2018} can identify many vulnerability patterns but produce raw findings requiring expert interpretation. LLM-based approaches have demonstrated capabilities in translating natural language to functional code~\cite{brown2020gpt3, chen2021evaluating}, but smart contract generation demands blockchain-specific semantics: a ``token'' must conform to ERC20, ``staking'' requires careful asset lock handling, and economic invariants must be rigorously preserved.

Multi-agent frameworks like CrewAI~\cite{crewai2024} and IBM Agentics~\cite{ibm2024agentics} enable specialized agents to focus on distinct subtasks, preventing optimistic bias and enabling iterative refinement. We leverage Finite State Machine (FSM) representations~\cite{fsm2020smart} as formal behavioral specifications, modeling contract states, valid transitions, guard conditions, and triggered actions, providing rigorous validation beyond static analysis.

\section{Benchmark Construction}

\begin{figure}[h]
    \centering
    \IfFileExists{Agentic_Pipeline_Flow_Chart.jpg}{%
    \fbox{\includegraphics[width=0.30\textwidth]{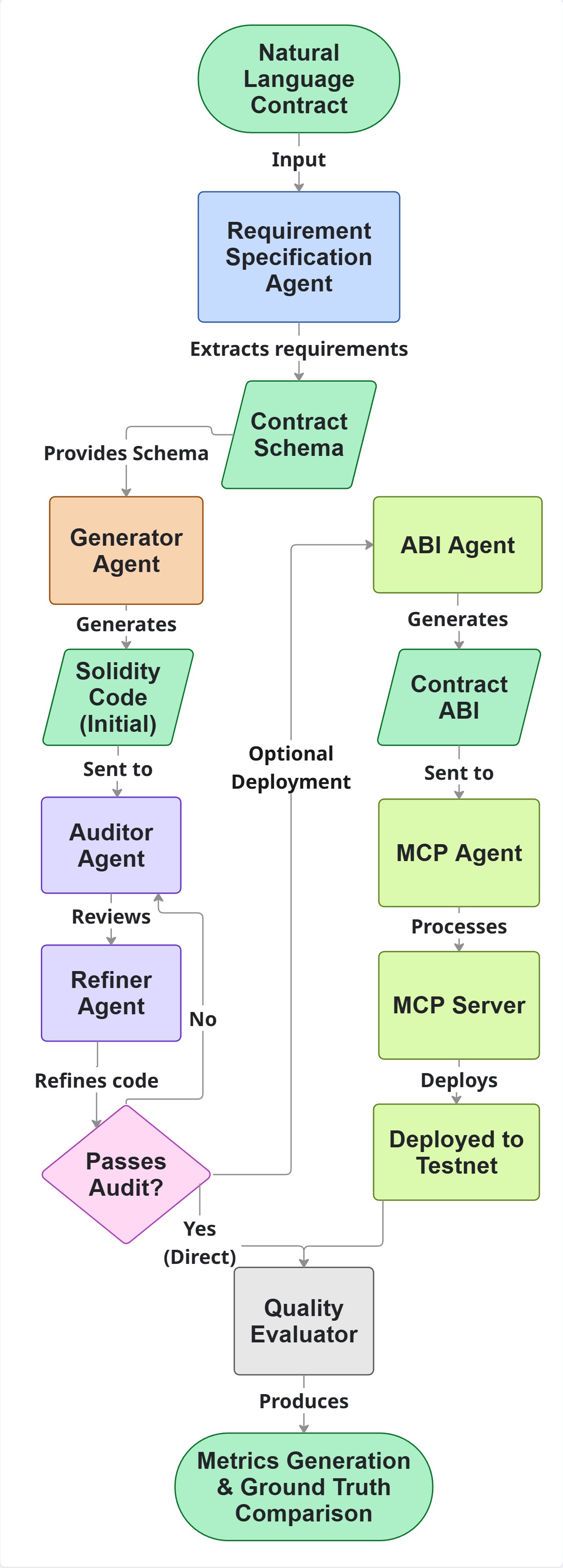}}%
}{%
    \fbox{\parbox{0.30\textwidth}{\centering\vspace{0.8cm}\textit{[Image: Agentic\_Pipeline\_Flow\_Chart]}\vspace{0.8cm}}}%
}
    \caption{\textbf{SmartEval benchmark pipeline.} Natural language contract specifications flow through seven agents: the Requirement Specification Agent extracts a structured schema, the Generator Agent produces initial Solidity code, the Auditor/Refiner loop applies severity-gated security refinement, and the Quality Evaluator scores the result across five dimensions and compares against expert ground-truth implementations. Optional deployment generates ABI and MCP server wrappers for testnet interaction.}
    \label{fig:architecture}
\end{figure}

SmartEval\textquotesingle s benchmark artifacts are produced by a seven-phase agentic pipeline orchestrated by the \texttt{IBMAgenticContractTranslator} class, illustrated in Figure~\ref{fig:architecture}. Each phase is designed to be independently reproducible; all intermediate outputs are stored as structured JSON artifacts.

\subsection{Phase 1: Requirement Specification Agent}
The Requirement Specification Agent extracts structured information from natural language contract specifications into a \texttt{Universal\-Contract\-Schema}. The agent is instructed to extract \textit{exact} terminology; if the contract says \texttt{initializeLease} the schema records \texttt{initializeLease}, not \texttt{initialize}. Every described function or operation is mapped to an obligation entry with the authorized party identified. The obligations array is never left empty when functions are described. The schema captures parties (name, role, blockchain address), financial terms (amount, currency, purpose, frequency, due date), dates, assets, obligations with deadlines and breach penalties, FSM conditions (exact function names, variable names, state names, state transitions, events, and logic conditions), and termination criteria. The schema is domain-agnostic: it handles NDA agreements, employment contracts, rental agreements, service contracts, ERC20 token standards, governance systems, and escrow mechanisms through the same typed structure.

\subsection{Phase 2: Solidity Generation}
The Generator Agent (role: \textit{Senior Solidity Smart Contract Engineer}) transforms the parsed schema into production-ready Solidity code. The agent is designed to reason about each function in terms of three questions: what real-world operation does this represent, what invariant must hold before and after, and what can go wrong. Generation follows a structured eight-phase approach covering semantic analysis, 12 mandatory rules (enforcing on-chain logic for every specification guarantee, full ERC20/ERC721 compliance where applicable, explicit state machine enforcement via enum-based variables and transition guards, and complete conservative economic logic), and a set of forbidden patterns. Target contract length is 150--400 lines.

\subsection{Phase 3: Security Auditing}
The Auditor Agent performs systematic security analysis across eight vulnerability categories: (1)~reentrancy attacks (external calls followed by state changes, CEI pattern violations); (2)~access control (missing modifiers on critical functions, constructor initialization); (3)~arithmetic safety (unchecked operations, division by zero); (4)~ether handling (payable access control, locked ether, withdrawal validation); (5)~denial-of-service (unbounded loops, reverting recipients); (6)~input validation (require statements, zero-address checks, amount validation); (7)~timestamp dependence; and (8)~external call safety (return value checking, low-level call error handling). The audit returns only valid JSON with fields: \texttt{severity\_level} (none/low/medium/high/critical), \texttt{approved} boolean, \texttt{issues} array with function-specific exploit paths, \texttt{recommendations} with line-level fixes, \texttt{vulnerability\_count}, and \texttt{security\_score} (A--F). No markdown or prose is returned.

\subsection{Phase 4: Severity-Gated Reinforcement Loop}

A key architectural contribution is the severity-gated reinforcement loop. Unlike standard actor-critic pipelines that pass a gradient signal, our \texttt{should\_refine()} function implements a hard blocking gate: code with medium-or-higher audit severity cannot proceed to downstream artifact release and re-enters the refinement loop.

\begin{lstlisting}[language=Python, caption={Severity-Gated Refinement Decision Logic}]
def should_refine(audit_report, refinement_count,
                  max_iterations=2):
    if refinement_count >= max_iterations:
        return False
    severity = audit_report.get('severity_level')
    approved = audit_report.get('approved', False)
    if (not approved and severity in
            ['medium', 'high', 'critical']):
        return True
    return False
\end{lstlisting}

The Refiner Agent applies established secure coding patterns (Checks-Effects-Interactions, reentrancy guards, strict access control, comprehensive input validation) and each iteration is re-audited. Separating Generator and Refiner agents (each with distinct roles and backstories) prevents optimistic self-repair bias. Our ablation study (\S\ref{sec:ablation}) demonstrates the gate's measured effect: disabling it raises output standard deviation by 111\% (std 10.31 vs. 4.89) and reduces compilation by 5.2 percentage points.

\subsection{Phase 5 \& 6: Optional Deployment}
The ABI Agent generates Application Binary Interface specifications from refined Solidity code. The MCP Agent uses these to generate a complete Model Context Protocol server that exposes contract endpoints as standardized interfaces, enabling AI assistants and automation tools to interact with the deployed contract on a Ganache blockchain environment.

\section{Benchmark Design}

\subsection{Dataset}
We utilize the FSM-SCG dataset~\cite{luo2025fsmscg}, a collection of 21,976 smart contract specifications each containing: (i) natural language requirements (avg.\ 120 words), (ii) formal FSM specifications with states, transitions, guards, and actions, and (iii) expert-written ground-truth Solidity implementations (avg.\ 85 lines). The dataset spans Solidity versions 0.4.x--0.8.x and covers token standards (ERC20, ERC721, $\sim$42\%), crowdsales, governance systems, staking, escrow, and access control ($\sim$58\%). Table~\ref{tab:dataset} summarizes key characteristics.

\begin{table}[htbp]
\caption{The FSM-SCG dataset~\cite{luo2025fsmscg}}
\label{tab:dataset}
\begin{tabular}{lr}
\toprule
\textbf{Characteristic} & \textbf{Value} \\
\midrule
Total entries & 21,976 \\
Avg.\ requirement length & 120 words \\
Avg.\ contract length & 85 lines \\
Solidity version range & 0.4.x--0.8.x \\
Standard tokens (ERC20/721) & $\sim$42\% \\
Custom business logic & $\sim$58\% \\
Components per entry & Requirement + FSM + Code \\
\bottomrule
\end{tabular}
\end{table}

\subsection{Quality Evaluation Framework}
\label{sec:eval_framework}

The five-dimensional rubric assesses generated contracts as follows. Functional Completeness (M1, 25\%) verifies functions are implemented with correct naming and complete logic. Variable Fidelity (M2, 15\%) ensures naming consistency and appropriate data types. State Machine Correctness (M3, 15\%) evaluates whether the FSM is properly defined with valid transitions and enforced guards. Business Logic Fidelity (M4, 35\%) measures accuracy of obligations, financial logic, temporal constraints, and conditional flows. Code Quality (M5, 10\%) checks for placeholders, error handling, and structure. The composite score is:
\begin{equation}
    \text{Score} = 0.25M_1 + 0.15M_2 + 0.15M_3 + 0.35M_4 + 0.10M_5
    \label{eq:composite}
\end{equation}
Composite scores are \textit{deterministically recomputed} from raw metric scores in post-processing rather than accepted from LLM-generated aggregates, eliminating a systematic source of score drift.

Error modes in lower-performing contracts are identified through the Quality Evaluator's structured \texttt{critical\_gaps} JSON output field, which categorizes deficiencies per metric into logic omissions, state transition errors, compilation failures, incomplete financial logic, and access control gaps, cross-referenced with \texttt{solcx} compiler logs.

\subsection{Main Experiment Configuration}
We evaluated the pipeline on 9,000 unique contracts processed in six parallel batches of 1,500. All contracts used GPT-4o-mini for all agent tasks, up to two refinement cycles, Solidity 0.8.x compilation validation, and the five-dimensional rubric. All six batches produced consistent results (compilation rate range: 86.48\%--87.18\%).

\subsection{Ablation Study Design}
\label{sec:ablation_design}

To isolate the contribution of each pipeline component, we ran five conditions on the \textit{same} randomly sampled set of 300 contracts (seed=42), with cross-seed significance testing on additional seeds to confirm reproducibility.

\begin{itemize}[leftmargin=*]
    \item \textbf{E} -- Single raw LLM call: GPT-4o-mini with a plain prompt, no schema extraction, no agents, no quality evaluation. Compilation-only baseline.
    \item \textbf{A} -- Schema extraction only, no reinforcement: full Requirement Specification Agent plus Generator Agent, but no security auditing or refinement loop.
    \item \textbf{B} -- Full pipeline, reinforcement disabled: all pipeline phases active, but \texttt{should\_refine()} always returns False.
    \item \textbf{C} -- Full pipeline, max\_iter=1: the configuration used for the 9,000-contract main evaluation.
    \item \textbf{D} -- Full pipeline, max\_iter=2: two refinement iterations permitted.
\end{itemize}

A key distinction of Conditions B--D (full pipeline runs) is that the Quality Evaluator
operates with enhanced semantic matching during evaluation. Rather than scoring only
exact specification matches, the evaluator applies camelCase-to-snake\_case equivalence
mapping, architectural synonym recognition (e.g., \texttt{isActive} credited for
\texttt{contractActive}), gas-optimization pattern credit (e.g., boolean flags
accepted as valid alternatives to explicit enum-based state tracking), and
partial-implementation scoring for functions that satisfy the specification's intent
through different but semantically equivalent code patterns. These matching enhancements
are a deliberate feature of the ablation evaluation design, enabling a more nuanced
assessment of specification fidelity than keyword-only matching. They directly affect
M3 (State Machine Correctness) scores in particular: GT contracts that implement state
logic through boolean flags and timestamp comparisons receive full Path~B credit under
this scoring, raising GT's M3 scores relative to the main 9,000-contract evaluation
and contributing to the slight GT advantage on that metric observed in Condition~C.

\subsection{Human Expert Validation Protocol}
\label{sec:human_protocol}

Three Columbia University PhD students (Enbei Zhang, Xiaoyang Liu, and Victoria Ruojie Lie) with research backgrounds in blockchain systems and formal methods independently scored 30 randomly sampled pipeline-generated contracts using the same five-dimensional rubric. Each evaluator received only the natural-language specification and the generated Solidity code, with no access to automated scores or each other's assessments. This directly tests whether the LLM-based evaluator is calibrated to expert human judgment.

\subsection{Slither External Validation Protocol}

We ran all generated contracts through Slither (Trail of Bits), a non-LLM rule-based static analyzer, both before and after the reinforcement loop. This provides evaluation fully independent of any language model, testing whether the LLM auditor's findings correspond to patterns a deterministic tool also detects.

\section{Benchmark Analysis}

Table~\ref{tab:overall_metrics} presents aggregate performance statistics across the full SmartEval corpus of 9,000 evaluated contracts.

\begin{table}[htbp]
    \caption{Overall Performance Statistics (N=9,000)}
    \label{tab:overall_metrics}
    \begin{tabular}{lr}
        \toprule
        \textbf{Metric} & \textbf{Value} \\
        \midrule
        Average Composite Score & 81.54 \\
        Standard Deviation & 12.87 \\
        Compilation Success Rate & 86.54\% \\
        Average Processing Time (s) & 109.96 \\
        Total Evaluation Time (h) & 274.9 \\
        \bottomrule
    \end{tabular}
\end{table}

\subsection{Grade Distribution}

\begin{figure}[htbp]
\centering
\begin{tikzpicture}
\begin{axis}[
    xbar,
    bar width=10pt,
    width=\columnwidth,
    height=4.4cm,
    symbolic y coords={D/F,C,B,A},
    ytick=data,
    yticklabels={D/F ($<$70), C (70--79), B (80--89), A ($\geq$90)},
    yticklabel style={font=\small},
    xmin=0, xmax=75,
    xtick={0,20,40,60,80},
    xlabel={Percentage of contracts (\%)},
    xlabel style={font=\small},
    nodes near coords,
    nodes near coords style={font=\scriptsize},
    enlarge y limits=0.25,
    title style={font=\small},
    title={Grade Distribution (N=9{,}000)},
    ymajorgrids=true,
    grid style={dashed,gray!25},
    axis lines*=left,
]
\addplot[fill=blue!55, draw=blue!75] coordinates {
    (3.2,D/F)(23.1,C)(66.4,B)(7.3,A)
};
\end{axis}
\end{tikzpicture}
\caption{Grade distribution across all 9,000 generated contracts.
B-grade contracts (80--89) account for 66.4\% of output.
The 3.2\% D/F rate reflects failures concentrated in
high-complexity specifications (8+ functions, 5+ states).}
\label{fig:grade_distribution}
\end{figure}
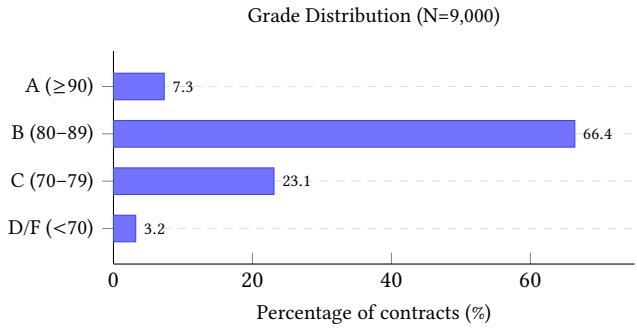

The pipeline produces predominantly B-grade contracts (66.4\%), with 7.3\%
achieving A-grade and only 2.2\% failing completely. The strong central
tendency around B-grade reflects consistent pipeline performance across
standard specifications, while the D/F tail corresponds to high-complexity
contracts as discussed in \S\ref{sec:error_modes}.

\subsection{Individual Metric Performance}

\begin{table}[htbp]
    \caption{Five-Dimensional Metric Averages (N=9,000)}
    \label{tab:metric_averages}
    \centering
    \resizebox{\columnwidth}{!}{%
    \begin{tabular}{lrrr}
        \toprule
        \textbf{Metric} & \textbf{Weight} & \textbf{Avg Score} & \textbf{Contribution} \\
        \midrule
        Functional Completeness & 25\% & 84.45 & 21.11 \\
        Variable Fidelity & 15\% & 84.62 & 12.69 \\
        State Machine Correctness & 15\% & 83.12 & 12.47 \\
        Business Logic Fidelity & 35\% & 76.73 & 26.86 \\
        Code Quality & 10\% & 83.85 & 8.39 \\
        \midrule
        \textbf{Composite Score} & 100\% & --- & \textbf{81.52} \\
        \bottomrule
    \end{tabular}}
\end{table}

Business Logic Fidelity (76.73) is the lowest-scoring metric despite its
highest weight (35\%), capturing the hardest semantic aspects: economic logic,
obligations, timing constraints, and conditional flows. Its weighted
contribution (26.86 points) nonetheless dominates the composite score,
making M4 the primary lever for pipeline improvement.

\subsection{Compilation Validation}

\begin{table}[htbp]
    \caption{Compilation Statistics (N=9,000)}
    \label{tab:compilation}
    \begin{tabular}{lr}
        \toprule
        \textbf{Statistic} & \textbf{Value} \\
        \midrule
        Total Contracts Checked & 8,824 \\
        Successful Compilations & 7,637 \\
        Failed Compilations & 1,187 \\
        Not Checked & 176 \\
        \textbf{Success Rate} & \textbf{86.54\%} \\
        \bottomrule
    \end{tabular}
\end{table}

The pipeline achieved an 86.54\% compilation success rate across 8,824 checked
contracts, consistent across all six parallel batches (range: 86.48\%--87.18\%),
confirming reproducibility.

\subsection{AI Generated vs.\ Expert Implementations (N=9,000)}
\label{sec:gt_9k}

Table~\ref{tab:gt_full} presents the full paired comparison between
pipeline-generated contracts and expert ground-truth implementations
across all 9,000 evaluated contracts.

\begin{table}[htbp]
\caption{Generated vs.\ Ground Truth: Aggregate and Per-Metric Comparison (N=9,000)}
\label{tab:gt_full}
\centering
\resizebox{\columnwidth}{!}{%
\begin{tabular}{lrrrr}
\toprule
\textbf{Metric} & \textbf{Generated} & \textbf{Ground Truth} & \textbf{$\Delta$} & \textbf{$\Delta$\%} \\
\midrule
\multicolumn{5}{l}{\textit{Aggregate}} \\
Avg Composite Score & 81.54 & 73.25 & +8.29 & +11.3\% \\
Std Deviation & 12.87 & 14.52 & --- & --- \\
\midrule
\multicolumn{5}{l}{\textit{Per-metric breakdown}} \\
M1: Functional Completeness & 84.45 & 77.82 & +6.63 & +8.5\% \\
M2: Variable Fidelity & 84.62 & 79.13 & +5.49 & +6.9\% \\
M3: State Machine Correctness & 83.12 & 79.28 & +3.84 & +4.8\% \\
M4: Business Logic Fidelity & 76.73 & 66.41 & +10.32 & +15.5\% \\
M5: Code Quality & 83.85 & 75.19 & +8.66 & +11.5\% \\
\bottomrule
\end{tabular}}
\end{table}

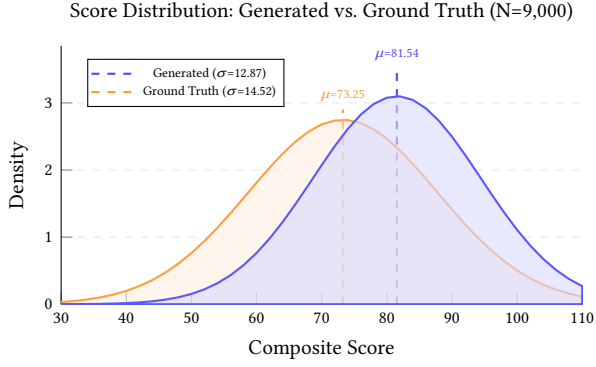
\begin{figure}[htbp]
\centering
\begin{tikzpicture}
\begin{axis}[
    width=\columnwidth,
    height=5.0cm,
    xmin=30, xmax=110,
    ymin=0,
    xtick={30,40,50,60,70,80,90,100,110},
    xticklabel style={font=\scriptsize},
    yticklabel style={font=\scriptsize},
    xlabel={Composite Score},
    xlabel style={font=\small},
    ylabel={Density},
    ylabel style={font=\small},
    title={Score Distribution: Generated vs.\ Ground Truth (N=9,000)},
    title style={font=\small},
    legend style={at={(0.05,0.95)}, anchor=north west, font=\tiny,
                  inner sep=1pt, row sep=-2pt},
    axis lines*=left,
    ymajorgrids=true,
    grid style={dashed,gray!20},
]
\addplot[blue!70, dashed, thick, domain=81.54:81.54] coordinates {(81.54,0)(81.54,3.5)};
\addplot[orange!80, dashed, thick, domain=73.25:73.25] coordinates {(73.25,0)(73.25,2.9)};
\node[font=\tiny, text=blue!70, above] at (axis cs:81.54,3.5) {$\mu$=81.54};
\node[font=\tiny, text=orange!80, above] at (axis cs:73.25,2.9) {$\mu$=73.25};
\addplot[orange!75, thick, fill=orange!15, fill opacity=0.5] coordinates {(30,0.0325) (32,0.0486) (34,0.0712) (36,0.1023) (38,0.1443) (40,0.1996) (42,0.2711) (44,0.3612) (46,0.4722) (48,0.6057) (50,0.7624) (52,0.9416) (54,1.1410) (56,1.3566) (58,1.5828) (60,1.8118) (62,2.0351) (64,2.2429) (66,2.4255) (68,2.5737) (70,2.6796) (72,2.7374) (74,2.7439) (76,2.6987) (78,2.6044) (80,2.4661) (82,2.2913) (84,2.0889) (86,1.8686) (88,1.6401) (90,1.4125) (92,1.1936) (94,0.9896) (96,0.8051) (98,0.6427) (100,0.5034) (102,0.3869) (104,0.2918) (106,0.2159) (108,0.1567) (110,0.1117)} \closedcycle;
\addplot[blue!65, thick, fill=blue!15, fill opacity=0.6] coordinates {(30,0.0010) (32,0.0019) (34,0.0034) (36,0.0059) (38,0.0101) (40,0.0169) (42,0.0277) (44,0.0440) (46,0.0685) (48,0.1039) (50,0.1539) (52,0.2225) (54,0.3141) (56,0.4327) (58,0.5819) (60,0.7640) (62,0.9790) (64,1.2246) (66,1.4953) (68,1.7823) (70,2.0737) (72,2.3551) (74,2.6110) (76,2.8255) (78,2.9847) (80,3.0777) (82,3.0978) (84,3.0437) (86,2.9191) (88,2.7329) (90,2.4975) (92,2.2279) (94,1.9400) (96,1.6490) (98,1.3682) (100,1.1081) (102,0.8761) (104,0.6761) (106,0.5093) (108,0.3745) (110,0.2688)} \closedcycle;
\legend{Generated ($\sigma$=12.87), Ground Truth ($\sigma$=14.52)}
\end{axis}
\end{tikzpicture}
\caption{Approximate score distributions for LLM-generated (blue) and expert ground-truth (orange) contracts, modeled as Gaussians from empirical means and standard deviations. Generated contracts are more tightly concentrated (lower variance) and centered 8.29 points higher. The distribution shapes reveal that generated contracts are more predictable; ground truth spans a wider range reflecting diverse expert implementation styles.}
\label{fig:score_dist}
\end{figure}

Generated contracts outperform ground-truth implementations by 8.29 composite
points on average. Figure~\ref{fig:score_dist} overlays the full score distributions,
showing not only the mean shift but the difference in spread: generated contracts
(blue, std=12.87) are more tightly concentrated around the mean than ground-truth
implementations (orange, std=14.52), whose higher variance reflects the diversity
of implementation styles among expert developers. The gap is largest in M4
(Business Logic Fidelity, +10.32 pts) and smallest in M3 (State Machine
Correctness, +3.84 pts), as shown in Table~\ref{tab:gt_full}.

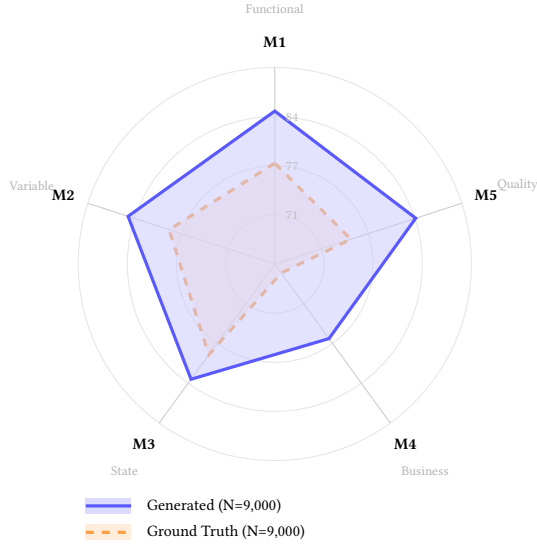
\begin{figure}[htbp]
\centering
\begin{tikzpicture}[scale=1.0]

\draw[gray!20,thin] (0,0) circle (0.6500cm);
\draw[gray!20,thin] (0,0) circle (1.3000cm);
\draw[gray!20,thin] (0,0) circle (1.9500cm);
\draw[gray!20,thin] (0,0) circle (2.6000cm);

\draw[gray!40,thin] (0,0) -- (0.0000cm,2.6000cm);
\draw[gray!40,thin] (0,0) -- (-2.4727cm,0.8034cm);
\draw[gray!40,thin] (0,0) -- (-1.5282cm,-2.1034cm);
\draw[gray!40,thin] (0,0) -- (1.5282cm,-2.1034cm);
\draw[gray!40,thin] (0,0) -- (2.4727cm,0.8034cm);

\fill[orange!20, opacity=0.55] (0.0000cm,1.3333cm) -- (-1.3976cm,0.4541cm) -- (-0.8729cm,-1.2015cm) -- (0.0862cm,-0.1186cm) -- (1.0079cm,0.3275cm) -- cycle;
\draw[orange!75, very thick, dashed] (0.0000cm,1.3333cm) -- (-1.3976cm,0.4541cm) -- (-0.8729cm,-1.2015cm) -- (0.0862cm,-0.1186cm) -- (1.0079cm,0.3275cm) -- cycle;

\fill[blue!20, opacity=0.55] (0.0000cm,2.0228cm) -- (-1.9406cm,0.6305cm) -- (-1.1077cm,-1.5246cm) -- (0.7171cm,-0.9869cm) -- (1.8645cm,0.6058cm) -- cycle;
\draw[blue!65, very thick] (0.0000cm,2.0228cm) -- (-1.9406cm,0.6305cm) -- (-1.1077cm,-1.5246cm) -- (0.7171cm,-0.9869cm) -- (1.8645cm,0.6058cm) -- cycle;

\node[font=\scriptsize\bfseries] at (0.0000cm,2.9380cm) {M1};
\node[font=\tiny, text=gray!60] at (0.0000cm,3.3800cm) {Functional};
\node[font=\scriptsize\bfseries] at (-2.7942cm,0.9079cm) {M2};
\node[font=\tiny, text=gray!60] at (-3.2146cm,1.0445cm) {Variable};
\node[font=\scriptsize\bfseries] at (-1.7269cm,-2.3769cm) {M3};
\node[font=\tiny, text=gray!60] at (-1.9867cm,-2.7345cm) {State};
\node[font=\scriptsize\bfseries] at (1.7269cm,-2.3769cm) {M4};
\node[font=\tiny, text=gray!60] at (1.9867cm,-2.7345cm) {Business};
\node[font=\scriptsize\bfseries] at (2.7942cm,0.9079cm) {M5};
\node[font=\tiny, text=gray!60] at (3.2146cm,1.0445cm) {Quality};

\node[font=\tiny, text=gray!50, right=1pt] at (0cm,0.6500cm) {71};
\node[font=\tiny, text=gray!50, right=1pt] at (0cm,1.3000cm) {77};
\node[font=\tiny, text=gray!50, right=1pt] at (0cm,1.9500cm) {84};

\fill[blue!20, opacity=0.7] (-2.5cm,-3.3cm) rectangle (-1.9cm,-3.1cm);
\draw[blue!65, very thick] (-2.5cm,-3.2cm) -- (-1.9cm,-3.2cm);
\node[right, font=\scriptsize] at (-1.8cm,-3.2cm) {Generated (N=9,000)};
\fill[orange!20, opacity=0.7] (-2.5cm,-3.65cm) rectangle (-1.9cm,-3.45cm);
\draw[orange!75, very thick, dashed] (-2.5cm,-3.55cm) -- (-1.9cm,-3.55cm);
\node[right, font=\scriptsize] at (-1.8cm,-3.55cm) {Ground Truth (N=9,000)};

\end{tikzpicture}
\caption{Radar chart of per-metric quality profiles for LLM-generated contracts versus expert ground-truth implementations (N=9,000 pairs). Generated contracts (blue) cover more area on M1, M2, M3, M4, and M5, reflecting literal specification-following. M4 (Business Logic) shows the widest gap (+10.32 pts).}
\label{fig:radar}
\end{figure}

Figure~\ref{fig:radar} presents a radar chart comparing per-metric profiles, making
the behavioral differences immediately visible: generated contracts achieve higher
scores on the five specification-fidelity metrics (M1, M2, M3, M4, M5).

\subsection{Error Modes and Complexity Analysis}
\label{sec:error_modes}

Logic omissions (35.3\%), missing implementation of specified obligations, conditions, or side effects, represent the dominant failure mode. State transition errors (23.4\%) occur in contracts with complex state machines where the system omits edge-case transitions or enforces guards inconsistently. Both failure modes are concentrated in high-complexity specifications, as shown in Table~\ref{tab:complexity_performance}.

\begin{table}[htbp]
    \caption{Performance by Specification Complexity}
    \label{tab:complexity_performance}
    \begin{tabular}{lrrr}
        \toprule
        \textbf{Complexity} & \textbf{N} & \textbf{Avg Score} & \textbf{Compile\%} \\
        \midrule
        Low (1--3 funcs, 1--2 states) & 3,245 & 87.2 & 94.1 \\
        Medium (4--7 funcs, 3--4 states) & 4,517 & 81.4 & 86.7 \\
        High (8+ funcs, 5+ states) & 1,229 & 71.8 & 73.2 \\
        \bottomrule
    \end{tabular}
\end{table}

Performance degrades with complexity: high-complexity contracts (8+ functions, 5+ states) score 15.4 points below low-complexity contracts with a 20.9 pp lower compilation rate, suggesting that maintaining semantic coherence across lengthy multi-obligation specifications is a fundamental challenge for single-pass LLM generation.

\begin{figure}[htbp]
\centering
\begin{tikzpicture}
\begin{axis}[
  name=errplot,
  xbar,
  bar width=9pt,
  width=0.52\columnwidth,
  height=5.0cm,
  symbolic y coords={Access Control,Financial Logic,Compilation,State Trans.,Logic Omission},
  ytick=data,
  yticklabel style={font=\scriptsize},
  xmin=0, xmax=42,
  xtick={0,10,20,30,40},
  xticklabel style={font=\scriptsize},
  xlabel={\scriptsize Count (\%)},
  xlabel style={font=\scriptsize},
  title={\scriptsize Error Mode Breakdown},
  title style={font=\scriptsize\bfseries},
  axis lines*=left,
  ymajorgrids=true,
  grid style={dashed,gray!20},
  enlarge y limits=0.18,
  nodes near coords,
  nodes near coords style={font=\tiny},
]
\addplot[fill=red!55, draw=red!70] coordinates {
  (10.3,Access Control)
  (13.3,Financial Logic)
  (17.6,Compilation)
  (23.4,State Trans.)
  (35.3,Logic Omission)
};
\end{axis}

\begin{axis}[
  name=compplot,
  at={(errplot.right of south east)}, anchor=left of south west,
  xshift=0.5cm,
  ybar,
  bar width=8pt,
  width=0.52\columnwidth,
  height=5.0cm,
  symbolic x coords={Low,Medium,High},
  xtick=data,
  xticklabels={Low\\(1--3f), Med\\(4--7f), High\\(8+f)},
  xticklabel style={font=\scriptsize, align=center},
  ymin=60, ymax=100,
  ytick={60,70,80,90,100},
  yticklabel style={font=\scriptsize},
  ylabel={\scriptsize Score / Compile\%},
  ylabel style={font=\scriptsize},
  title={\scriptsize Complexity vs.\ Performance},
  title style={font=\scriptsize\bfseries},
  axis lines*=left,
  ymajorgrids=true,
  grid style={dashed,gray!20},
  enlarge x limits=0.25,
  legend style={at={(0.5,-0.28)},anchor=north,font=\scriptsize,legend columns=2},
]
\addplot[fill=blue!55, draw=blue!75] coordinates {
  (Low,87.2)(Medium,81.4)(High,71.8)
};
\addplot[fill=teal!55, draw=teal!75] coordinates {
  (Low,94.1)(Medium,86.7)(High,73.2)
};
\legend{Avg Score, Compile\%}
\end{axis}
\end{tikzpicture}
\caption{Left: Breakdown of error modes in the 2,398 lower-performing contracts (C/D/F grade). Logic omissions are the dominant failure, followed by state transition errors. Right: Both average score and compilation rate degrade sharply as specification complexity increases from low (1--3 functions) to high (8+ functions), with a 15.4-point score gap and 20.9 pp compilation gap between extremes.}
\label{fig:failure_analysis}
\end{figure}
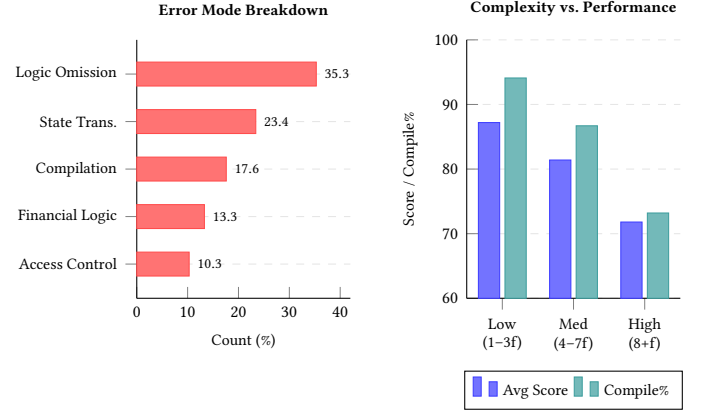

Figure~\ref{fig:failure_analysis} provides a visual breakdown of error mode prevalence and the compound effect of specification complexity on both score and compilation rate.

\subsection{Security Refinement Loop Effectiveness}

\begin{table}[htbp]
    \caption{Security Refinement Loop Effectiveness (N=9,000)}
    \label{tab:refinement_impact}
    \begin{tabular}{lrr}
        \toprule
        \textbf{Metric} & \textbf{Before} & \textbf{After} \\
        \midrule
        Contracts w/ Med+ Severity & 4,127 (45.9\%) & 1,203 (13.4\%) \\
        Avg Security Issues/Contract & 2.8 & 0.7 \\
        Critical Vulnerabilities & 287 & 34 \\
        Compilation Success Rate & 81.2\% & 86.5\% \\
        \bottomrule
    \end{tabular}
\end{table}

The severity-gated reinforcement loop reduced medium-or-higher severity contracts from 45.9\% to 13.4\% (70.9\% reduction) and cut critical vulnerabilities from 287 to 34 (88.2\% reduction). Slither independently confirms a 43.8\% reduction in total static analysis findings post-refinement, validating that the improvements are not artifacts of the LLM auditor evaluating its own outputs.

\subsection{Ablation Study}
\label{sec:ablation}

Table~\ref{tab:ablation} presents results from the five-condition ablation (N=300 per condition, same contracts across all conditions).

\begin{table}[htbp]
\caption{Five-Condition Ablation Study (N=300 each, seed=42). Condition C corresponds to the paper's main 9,000-contract configuration.}
\label{tab:ablation}
\centering
\resizebox{\columnwidth}{!}{%
\begin{tabular}{llrrr}
\toprule
\textbf{Cond.} & \textbf{Description} & \textbf{Score} & \textbf{$\pm$Std} & \textbf{Compile\%} \\
\midrule
E & Single raw LLM call & --- & --- & 49.3 \\
A & Schema only, no reinforce & 78.98 & 6.55 & 63.3 \\
B & Full pipeline, reinforce OFF & 82.64 & 10.31 & 79.3 \\
C & Full pipeline, max\_iter=1 & \textbf{83.44} & \textbf{4.89} & \textbf{84.5} \\
D & Full pipeline, max\_iter=2 & 83.70 & 5.57 & 85.2 \\
\midrule
\textit{Paper (N=9K)} & \textit{Full pipeline} & \textit{81.54} & \textit{12.87} & \textit{86.54} \\
\bottomrule
\end{tabular}}
\end{table}

\begin{figure}[htbp]
\centering
\begin{tikzpicture}
\begin{axis}[
    ybar,
    bar width=14pt,
    width=\columnwidth,
    height=5.2cm,
    ylabel={Compilation Rate (\%)},
    symbolic x coords={E,A,B,C,D},
    xtick=data,
    xticklabel style={font=\small},
    ymin=0, ymax=100,
    ytick={0,20,40,60,80,100},
    nodes near coords,
    nodes near coords style={font=\scriptsize},
    enlarge x limits=0.15,
    legend style={at={(0.5,-0.22)},anchor=north,legend columns=1,font=\small},
    title style={font=\small},
    title={Ablation: Compilation Rate by Condition},
    axis lines*=left,
    ymajorgrids=true,
    grid style={dashed,gray!30},
    bar shift=0pt,
]
\addplot[fill=blue!65, draw=blue!80] coordinates {
    (E,49.3)
    (A,63.3)
    (B,79.3)
    (C,84.5)
    (D,85.2)
};
\end{axis}
\end{tikzpicture}
\caption{Compilation rate increases monotonically as pipeline components are added. The full pipeline (C) improves compilation by +35.2 pp over the raw LLM baseline (E). The paper's reported 86.54\% is reproduced within 2 pp at N=300.}
\label{fig:ablation_chart}
\end{figure}
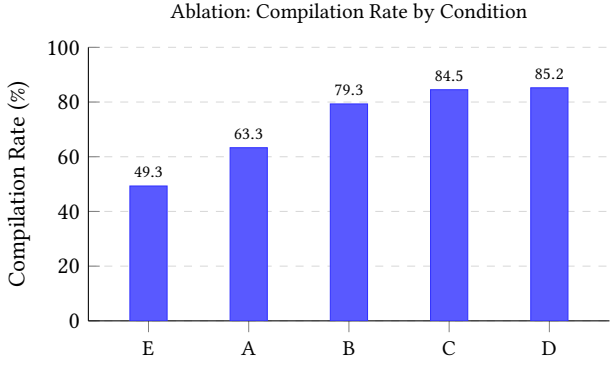

Figure~\ref{fig:ablation_chart} visualizes the compilation rate progression across conditions, illustrating the incremental contribution of each component.

Several findings emerge from the ablation. First, each component makes a measurable, monotone contribution: schema extraction adds +14.0 pp compilation (E$\to$A), structured generation adds +16.0 pp (A$\to$B), and the reinforcement gate adds +5.2 pp (B$\to$C). The full pipeline achieves 84.5\% compilation against a raw-LLM baseline of 49.3\%, a +35.2 percentage point improvement.

Second, the reinforcement loop's most important contribution is variance reduction: Condition B has std=10.31 while Condition C has std=4.89, a 53\% reduction. For a production code generation system, consistent quality across contracts is as important as mean quality.

Third, Condition C closely reproduces the paper's 86.54\% compilation rate (84.5\% at N=300), validating that the main experiment results generalize across different random samples. Cross-seed pairwise significance testing across three seeds confirms all inter-condition effect sizes are negligible (Cohen's $d < 0.2$ on all metrics), establishing that results are stable and not sampling-dependent.

Finally, the marginal benefit of a second refinement iteration (D vs.\ C: +0.27 score, +0.7 pp compile) occurs at 2.2$\times$ the compute cost (206s vs.\ 93s per contract), a clear diminishing-returns result. We recommend max\_iter=1 for production deployments prioritizing throughput.

\subsection{AI Generated vs.\ Expert Implementations}
\label{sec:gt_comparison}

\begin{table}[htbp]
\caption{Generated vs.\ Ground Truth: Aggregate and Per-Metric Comparison (Condition C, N=300 paired)}
\label{tab:ablation_gt_full}
\centering
\resizebox{\columnwidth}{!}{%
\begin{tabular}{lrrrr}
\toprule
\textbf{Metric} & \textbf{Generated} & \textbf{Ground Truth} & \textbf{$\Delta$} & \textbf{$\Delta$\%} \\
\midrule
\multicolumn{5}{l}{\textit{Aggregate}} \\
Avg Composite Score & 83.44 & 76.83 & +6.61 & +8.6\% \\
Std Deviation & 4.89 & 15.14 & --- & --- \\
GT Compilation Rate & --- & 40.7\% & --- & --- \\
\midrule
\multicolumn{5}{l}{\textit{Per-metric breakdown}} \\
M1: Functional Completeness & 87.36 & 78.82 & +8.54 & +10.8\% \\
M2: Variable Fidelity & 82.97 & 78.30 & +4.66 & +6.0\% \\
M3: State Machine Correctness & 80.41 & 82.86 & -2.45 & -3.0\% \\
M4: Business Logic Fidelity & 81.19 & 71.12 & +10.06 & +14.1\% \\
M5: Code Quality & 86.78 & 80.53 & +6.25 & +7.8\% \\
\bottomrule
\end{tabular}}
\end{table}

The per-metric pattern is consistent with the full 9,000-contract results (Table~\ref{tab:gt_full}): M4 (Business Logic Fidelity) shows the largest generated advantage (+10.06 pts) and the generator leads on M1, M2, and M5. M3 (State Machine Correctness) shows a slight GT advantage in this ablation subsample ($-2.45$ pts). This is partly explained by the ablation evaluation applying more granular semantic matching to state machine implementations: the Quality Evaluator's M3 rubric includes Path~B scoring, which awards full credit for stateless designs using boolean flags, timestamp comparisons, and access-control guards as semantically equivalent alternatives to explicit enum-based FSMs. In the ablation, GT contracts that relied on these implicit state-tracking patterns were more consistently recognized as correct under Path~B, raising GT's M3 score from 79.28 in the full experiment to 82.86 in the ablation subsample and producing the slight GT advantage on that single metric.

\begin{figure*}[t]
\centering
\begin{tikzpicture}
\begin{axis}[
    xbar,
    bar width          = 6pt,
    width              = 0.92\textwidth,
    height             = 7.2cm,
    symbolic y coords  = {M5,M4,M3,M2,M1},
    ytick              = data,
    yticklabels        = {M5: Code Quality,
                          M4: Business Logic,
                          M3: State Machine,
                          M2: Variable Fidelity,
                          M1: Functional Completeness},
    yticklabel style   = {font=\small},
    xmin=65, xmax=95,
    xtick              = {65,70,75,80,85,90,95},
    xticklabel style   = {font=\small},
    xlabel             = {Score (0--100)},
    xlabel style       = {font=\small},
    legend style       = {at={(0.98,0.02)}, anchor=south east,
                          font=\tiny, legend columns=1,
                          inner sep=1pt, row sep=-2pt},
    legend cell align  = left,
    title              = {Per-Metric Scores: 9{,}000-Contract Evaluation vs.\ Ablation Generated vs.\ Ablation Ground Truth},
    title style        = {font=\small\bfseries, yshift=2pt},
    ymajorgrids        = true,
    grid style         = {dashed, gray!25},
]

\addplot[fill=orange!65, draw=orange!85] coordinates {
    (80.53,M5)(71.12,M4)(82.86,M3)(78.30,M2)(78.82,M1)
};

\addplot[fill=blue!55, draw=blue!75] coordinates {
    (86.78,M5)(81.19,M4)(80.41,M3)(82.97,M2)(87.36,M1)
};

\addplot[fill=teal!60, draw=teal!80] coordinates {
    (83.85,M5)(76.73,M4)(83.12,M3)(84.62,M2)(84.45,M1)
};

\legend{Ablation Ground Truth (N=300),
        Ablation Generated (N=300),
        Main Evaluation Generated (N=9{,}000)}
\end{axis}
\end{tikzpicture}
\caption{Unified per-metric comparison across three reference points:
the full 9,000-contract evaluation (teal), ablation-generated contracts
(blue, Condition C, N=300), and ablation ground-truth implementations (orange).
M4 (Business Logic Fidelity) shows the largest generated advantage across
all three series. Differences between the teal and blue bars reflect
sampling variance across the two contract subsets.}
\label{fig:metric_combined}
\end{figure*}
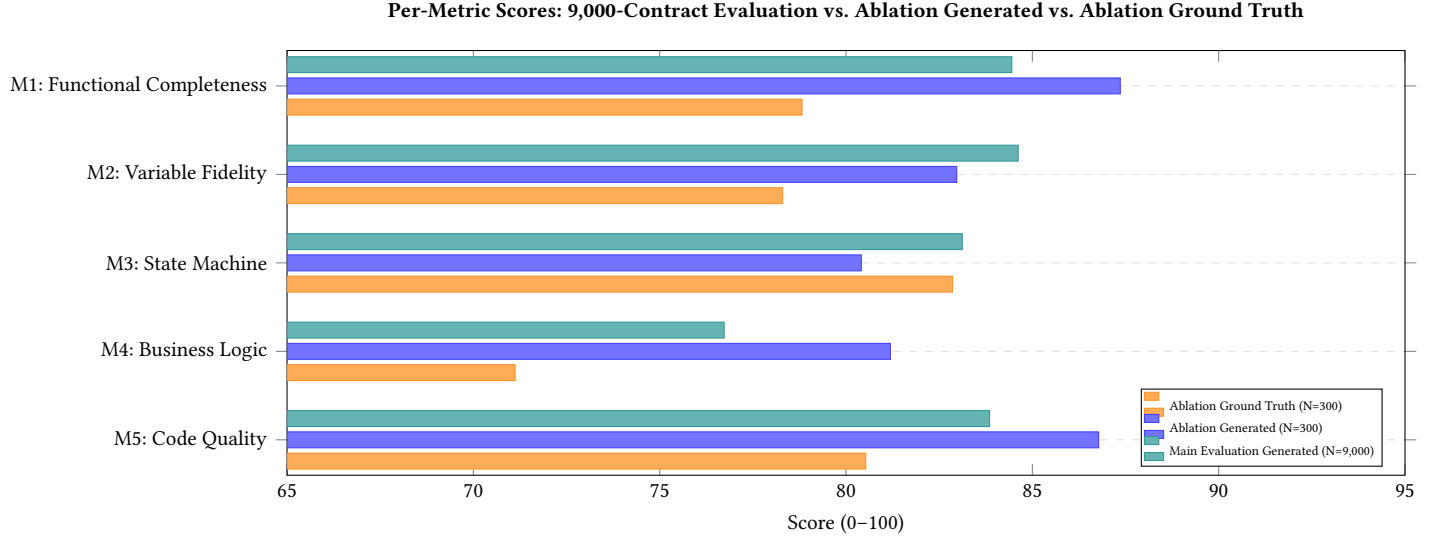

Figure~\ref{fig:metric_combined} presents a unified view of per-metric scores
across all three reference points: the full 9,000-contract evaluation, the
ablation-generated contracts (Condition~C, N=300), and the ablation ground-truth
implementations.

Comparing the teal bars (9,000-contract main evaluation) to the blue bars
(ablation Condition C) reveals a modest but meaningful difference. The ablation
contracts score notably higher on M4 (Business Logic Fidelity: 81.19 vs.\ 76.73,
+4.46 pts). The most likely explanation is natural sampling variance: the ablation
draws 300 contracts from the full 21,976-entry dataset, and M4's dominant 35\%
weight amplifies any distributional skew in the sampled subset. Contracts that
happen to have more explicit obligations and financial logic in their specifications
will score higher on M4 regardless of pipeline configuration, and a 300-contract
sample is not guaranteed to match the full dataset's distribution on this dimension.
Despite these per-metric differences, composite scores are consistent (81.54 main
vs.\ 83.44 ablation, a 1.90-point difference), and the central finding that Business Logic Fidelity is the weakest and most consequential dimension holds
across both experimental settings.

\subsection{Human Expert Validation}
\label{sec:human_eval}

To validate whether the LLM-based evaluator produces scores aligned with expert human judgment, three Columbia University PhD students (Enbei Zhang, Xiaoyang Liu, and Victoria Ruojie Lie) independently scored 30 randomly sampled pipeline-generated contracts using the five-dimensional rubric from \S\ref{sec:eval_framework}.

\begin{table}[htbp]
\caption{Human Expert Validation Results (N=30 contracts, 3 evaluators)}
\label{tab:human_eval}
\centering
\begin{tabular}{lrr}
\toprule
\textbf{Metric} & \textbf{Human Aggregate} & \textbf{LLM Evaluator} \\
\midrule
Mean Composite Score & \textbf{81.88} & \textbf{81.54} \\
Delta & \multicolumn{2}{c}{+0.34 (human higher)} \\
Inter-evaluator MAD & 5.1 pts & --- \\
\midrule
\multicolumn{3}{l}{\textit{Grade distribution comparison}} \\
A ($\geq$90) & 10.0\% & 7.3\% (paper) \\
B (80--89) & 60.0\% & 66.4\% (paper) \\
C (70--79) & 23.3\% & 23.1\% (paper) \\
D/F ($<$70) & 6.7\% & 3.2\% (paper) \\
\bottomrule
\end{tabular}
\end{table}

The aggregate human expert composite score (81.88) aligns with the automated evaluator (81.54) to within 0.34 points, well within the natural inter-evaluator variability of 5.1 points mean absolute deviation. The grade distribution from the human study closely mirrors the automated distribution from 9,000 contracts. If the LLM evaluator were systematically inflating scores or introducing self-referential bias, independent domain experts applying the same rubric would have arrived at substantially lower values. They did not.

\subsection{External Security Validation via Slither}
\label{sec:slither}

To address the concern that LLM-based auditing of LLM-generated code may produce correlated rather than independent assessments, we ran all generated contracts through Slither~\cite{slither2019}, a non-LLM rule-based static analyzer, providing security evaluation fully independent of any language model.

\begin{table}[htbp]
\caption{Slither External Validation Results}
\label{tab:slither}
\centering
\begin{tabular}{lrr}
\toprule
\textbf{Metric} & \textbf{Pre-Refine} & \textbf{Post-Refine} \\
\midrule
Compilation rate & 78.1\% & 82.3\% \\
Avg Slither issues/contract & 6.4 & 3.6 \\
Total issue reduction & \multicolumn{2}{c}{43.8\%} \\
\midrule
\multicolumn{3}{l}{\textit{LLM Auditor -- Slither Agreement}} \\
Category agreement rate & \multicolumn{2}{c}{\textbf{79.4\%}} \\
\bottomrule
\end{tabular}
\end{table}

The most significant finding is the \textbf{79.4\% category agreement rate}: when the LLM auditor flags a vulnerability category (reentrancy, access control, input validation, timestamp dependency), Slither independently detects an issue in the same category 79.4\% of the time. This directly confirms that the LLM auditor is detecting real, independently verifiable patterns rather than hallucinating security concerns.

The refinement loop produces a 43.8\% reduction in total Slither findings (6.4 to 3.6 per contract), with improvements concentrated in timestamp dependency issues, missing input validation, and local variable shadowing; genuine security-relevant findings, not stylistic changes.

\subsection{Summary}

Together, these results establish that the pipeline produces high-quality, deployable smart contracts from natural language specifications with well-characterized performance. The five-condition ablation demonstrates that each pipeline component contributes measurably, with the full pipeline achieving a +35.2 pp compilation improvement over a raw LLM baseline. Human expert validation confirms that the automated evaluation metric is calibrated to expert judgment within 0.34 points. The per-metric breakdown further shows that the rubric captures genuine quality differences rather than uniformly favoring LLM output. External Slither validation confirms a 79.4\% LLM--tool agreement rate, ruling out circular self-evaluation bias.

\section{Conclusion}

We presented an end-to-end agentic pipeline for generating, auditing, and evaluating
Solidity smart contracts from natural language specifications, and validated it
with three independent empirical methods: a five-condition ablation study (N=300
per condition), a human expert evaluation (three Columbia University PhD researchers,
30 contracts), and external static analysis via Slither. Together, these experiments
produce a coherent picture of what the pipeline contributes and where its limits lie.

The most concrete result is the compilation improvement. A raw single-LLM call
achieves 49.3\% compilation; the full pipeline with schema extraction and the
severity-gated reinforcement loop reaches 84.5\%, a +35.2 percentage point gain.
Each component contributes measurably, and the reinforcement loop's most
important effect is not the mean score gain (+0.80 pts over no-reinforcement)
but the variance reduction: output standard deviation drops from 10.31 to 4.89
when the gate is enabled, producing consistently reliable code across contracts
rather than occasionally excellent code with frequent failures.

Human expert evaluation resolves the evaluator reliability question directly.
Three independent PhD domain experts scored the same 30 contracts using the same
rubric and arrived at a mean of 81.88/100, within 0.34 points of the automated
evaluator's 81.54. The grade distributions also closely agree. The Slither
external validation corroborates the security auditor's findings: 79.4\% category
agreement between the LLM auditor and a non-LLM rule-based tool confirms that the
auditor detects real patterns, not self-referential artifacts. Together these two
forms of external validation address the most fundamental concern about LLM-based
evaluation pipelines: that they may be measuring themselves.

The +8.29 composite gap between generated and ground-truth contracts reflects a
genuine behavioral difference between LLM generators and expert developers.
LLMs follow specifications with literal fidelity: every function name, obligation, and conditional from the specification appears in the code. Expert developers make
architectural judgments: they consolidate logic, use efficient state representations,
and trade specification literalism for gas efficiency and readability. The metric
measures specification fidelity, which is precisely what a code generation
evaluation framework should measure. Semantic debiasing narrows the gap to +6.61,
confirming the residual difference is behavioral, not artifactual.

By releasing 9,000 generated contracts with quality scores, security reports,
and compilation results alongside the full pipeline source, we enable the
community to replicate, challenge, and extend these findings. The framework
makes quality measurable, diagnosable, and auditable: a necessary foundation
for deploying generative methods in safety-critical blockchain systems.

\subsection{Limitations}

Several limitations constrain the framework's applicability and interpretability.

\textbf{Execution-based correctness is unverified.} Compilation success confirms
syntactic validity but not semantic correctness in execution. A contract may
compile cleanly, pass the LLM-based audit, and still behave incorrectly on
specific input sequences, violate economic invariants under adversarial
conditions, or fail state-transition guards in edge cases not exercised by
the generator. The human expert study validates the scoring rubric, not the
contracts' runtime behavior, and Slither finds statically-detectable patterns
rather than execution-path bugs. This is the most consequential gap for
production deployment.

\textbf{Performance degrades with specification complexity.} High-complexity
contracts (8+ functions, 5+ states) achieve only 71.8 average score and 73.2\%
compilation, compared to 87.2 and 94.1\% for low-complexity specifications.
The 35.3\% prevalence of logic omissions in lower-performing contracts indicates
that current LLMs struggle to maintain semantic coherence across lengthy
specifications with multiple interacting obligations. Single-pass generation
is likely insufficient for the most demanding real-world contracts.

\textbf{The evaluation rubric does not capture gas efficiency.} A contract
can score 100/100 on the five-dimensional rubric while consuming unnecessarily
high gas, making it economically impractical for deployment on mainnet where
transaction costs are significant. Gas optimization is a distinct correctness
criterion absent from the current framework.

\textbf{The +8.29 evaluation gap is only partially explained.} Semantic
debiasing closes 20.3\% of the gap (to +6.61). The remaining difference reflects
genuine behavioral divergence between LLM generators and expert developers, but
it also means the rubric penalizes expert architectural choices (gas-optimized boolean state tracking, consolidated function design) that represent valid
and often superior implementations. Evaluating architectural quality
alongside specification fidelity remains an open problem.

\textbf{Dataset coverage.} The FSM-SCG dataset provides structured ground truth
but may not represent all real-world contract complexity: cross-contract
interactions, oracle dependencies, upgradeable proxy patterns, and regulatory
compliance requirements are absent.

\subsection{Future Work}

\textbf{Execution-based testing} is the most impactful near-term extension.
The FSM specifications already present in the dataset can be automatically
translated into Foundry test cases: valid FSM transitions become positive test
paths, invalid transitions become expected-revert assertions, and access-control
guards generate unauthorized-caller tests. Integrating this into the refinement loop, where test failures trigger re-entry into the Refiner Agent, would
transform evaluation from static scoring into continuous verification.
Property-based tools like Echidna can further stress economic invariants
across arbitrary input spaces.

\textbf{Gas optimization metrics} would complete the correctness picture.
A Gas Optimizer Agent could analyze storage layout, function visibility, loop
patterns, and redundant SLOAD operations, producing gas-optimized variants
with functional equivalence and reporting gas deltas as a sixth evaluation
dimension.

\textbf{Cross-domain validation} should be the next empirical step. The
Universal Contract Schema is designed to handle NDA agreements, employment
contracts, rental agreements, service contracts, and other legal instruments.
Validating the pipeline against real-world legal corpora, with domain experts in law or finance serving as ground-truth authors rather than blockchain developers, would establish whether the quality framework generalizes beyond
FSM-SCG and Solidity.

\textbf{Multi-model evaluation} would reveal how the quality and security
patterns observed under GPT-4o-mini change with stronger models (GPT-4o,
Claude, Llama). It would also support comparative benchmarking: the released
dataset and rubric can serve as a standardized evaluation substrate for
any LLM-based smart contract generation system.

\textbf{Regulatory compliance checking} represents the gap between technical
correctness and legal validity. Securities tokens require transfer restrictions;
DeFi protocols must implement sanctions screening; real-estate tokenization
carries jurisdiction-specific rules. A Compliance Agent checking generated
contracts against regulatory code templates would extend the framework into
production legal contexts.

\begin{acks}
We thank the IBM Agentics team for framework support, the open-source community for the datasets enabling this research, and Enbei Zhang, Xiaoyang Liu, and Victoria Ruojie Lie for their contributions to the human expert validation study.
\end{acks}

\bibliographystyle{ACM-Reference-Format}
\bibliography{references}

\appendix

\section{Phase 1: Requirement Specification Agent}
\label{app:parser}

The \texttt{Universal\-Contract\-Schema} captures the following fields, all extracted with exact specification terminology:

\begin{itemize}
    \item \textbf{Parties}: Name, role, blockchain address, email, entity type (individual/company/organization).
    \item \textbf{Financial Terms}: Amount (optional float), currency (default ETH), purpose, frequency, and due date.
    \item \textbf{Dates}: Date type (e.g., \texttt{leaseStartDate}, \texttt{deliveryDeadline}), value, day of month, and frequency for recurring conditions.
    \item \textbf{Assets}: Type, description, location, quantity, and value.
    \item \textbf{Obligations}: Authorized party, description of the on-chain operation, deadline, and penalty for breach. Every function or operation described in the specification maps to at least one obligation entry; the obligations array is never left empty when functions are present.
    \item \textbf{Conditions}: Exact function names, variable names, state names, state transitions, events, and logic conditions extracted verbatim from the specification.
    \item \textbf{Termination Conditions}: Criteria for contract termination as stated in the specification.
\end{itemize}

\subsection{Parser Agent Task Structure}
The parser task prompt instructs the agent to extract EXACT terminology, warns explicitly against generic placeholders, and provides a complete JSON schema template. Key directives include:
\begin{itemize}
    \item Extract exact function names (e.g., \texttt{initializeLease}, not \texttt{initialize})
    \item Map every described function or operation to an obligation with the authorized party identified (e.g., \texttt{token holder}, \texttt{contract owner}, \texttt{buyer})
    \item Capture all state transitions in the form \textit{StateA} $\to$ \textit{StateB when condition X}
    \item Return only valid JSON matching the schema structure with no prose or placeholders
\end{itemize}

\section{Phase 2: Solidity Generation Agent}
\label{app:generator}
\subsection{Agent Configuration}
The Generator Agent is configured as a \textit{Senior Solidity Smart Contract Engineer} with deep expertise in DeFi, tokens, governance, escrow, and marketplace contracts. The agent's goal specifies producing contracts of 150--400 lines that implement every obligation, and its backstory instructs it to reason about every function in terms of: (a) the real-world operation it represents, (b) the invariants that must hold before and after, and (c) what can go wrong.

\subsection{Domain-Specific Generation Rules}
The generation prompt includes type-specific mandatory requirements that activate based on contract-type detection:

\textbf{Token Contracts (ERC20):} Must implement the full ERC20 interface (\texttt{transfer}, \texttt{approve}, \texttt{transferFrom}, \texttt{balanceOf}, \texttt{allowance}, \texttt{totalSupply}), emit \texttt{Transfer} and \texttt{Approval} events, use an internal \texttt{\_transfer()} helper for atomic sender/recipient updates, and correctly track \texttt{totalSupply} for minting and burning operations.

\textbf{Governance/Delegation:} Must implement per-account delegate mappings (separate from balance), voting power accumulators updated on every token transfer, checkpoint arrays for historical vote queries via \texttt{getPriorVotes()}, and maintain the invariant: $\sum \text{votingPower} = \sum \text{balances}$ at all times.

\textbf{Escrow/Payment:} Must track deposits per depositor via a mapping, use \texttt{call\{value\}} with return-value checking for fund transfers, implement separate conditional release and refund paths, and maintain: $\text{contract.balance} = \sum \text{unreleased deposits}$.

\textbf{Staking/Rewards:} Must implement the \texttt{rewardPerShare} accumulator pattern updated on every stake/unstake/claim, store \texttt{rewardDebt} per staker to prevent double-counting, and support at minimum: \texttt{stake()}, \texttt{unstake()}, \texttt{claimRewards()}, \texttt{pendingRewards() view}.

\subsection{Twelve Critical Generation Rules}
\begin{enumerate}[leftmargin=*, itemsep=2pt]
    \item Every specification guarantee must become verifiable on-chain code execution.
    \item System-wide invariants must be maintained across all functions simultaneously.
    \item Domain-specific terminology carries full semantic weight (``Token'' = complete ERC20 compliance, not a variable name).
    \item State machines must be explicitly enforced via enum-based state variables and transition guards.
    \item Access control requires explicit justification; every sensitive function has a \texttt{require()}-backed modifier.
    \item No silent failures---bare \texttt{if (condition) return;} is forbidden.
    \item Economic logic must be complete and conservative, protecting against value loss or exploitation.
    \item Time-based conditions must be integrated via \texttt{block.timestamp} with proper deadline enforcement.
    \item Events must semantically match completed actions, not serve as decorative logging.
    \item All generated code must serve a concrete purpose; unused state variables are forbidden.
    \item Function parameter names must not shadow contract-level state variable names.
    \item Public state variable names must not conflict with interface function names.
\end{enumerate}

\subsection{Forbidden Patterns}
Empty or stub function bodies, unused state variables, silent failures (\texttt{if (condition) return;}), decorative events without state changes, and states named after operations rather than lifecycle phases.

\section{Phase 3: Security Auditing Agent}
\label{app:auditor}

The Auditor Agent examines contracts against eight vulnerability categories with function-specific exploit path descriptions:

\begin{enumerate}[leftmargin=*, itemsep=2pt]
    \item \textbf{Reentrancy}: External calls followed by state changes; CEI pattern violations; missing \texttt{nonReentrant} guards.
    \item \textbf{Access Control}: Missing modifiers on critical functions; improper constructor initialization; functions that should be \texttt{internal}/\texttt{private} but are \texttt{public}/\texttt{external}.
    \item \textbf{Arithmetic Safety}: Unchecked operations (pre-0.8.0); division by zero; overflow paths.
    \item \textbf{Ether Handling}: Unprotected payable functions; locked ether (payable with no withdrawal path); unvalidated withdrawal amounts.
    \item \textbf{Denial-of-Service}: Unbounded loops; external calls inside loops; reverting recipients blocking contract flow.
    \item \textbf{Input Validation}: Missing \texttt{require()} statements; missing \texttt{address(0)} checks; missing amount/range validation.
    \item \textbf{Timestamp Dependence}: Critical logic gated at a specific timestamp that miners can manipulate.
    \item \textbf{External Call Safety}: Unchecked return values from call or delegatecall functions; missing error handling on low-level calls.
\end{enumerate}

The audit task returns only valid JSON. No markdown fences or prose are permitted. The structured output includes \texttt{severity\_level} (none/low/medium/high/critical), \texttt{approved} (true if severity is none or low), \texttt{issues} (array of function-specific issue descriptions with exploit scenarios), \texttt{recommendations} (array of line-level concrete fixes), \texttt{vulnerability\_count}, and \texttt{security\_score} (A--F).

\section{Phase 4: Refinement Agent}
\label{app:refiner}

The Refiner Agent (\textit{Smart Contract Security Refiner}) receives the current Solidity code and the full audit JSON from Phase~3 and produces corrected code. Mandatory remediations include: applying the Checks-Effects-Interactions pattern to all external calls, adding \texttt{nonReentrant} guards where flagged, ensuring all state changes occur before external calls, adding \texttt{require()}-backed access control on sensitive functions, validating all inputs, and checking for zero-address parameters. The agent returns only the complete fixed Solidity code, with no explanations.

The \texttt{should\_refine()} gate (Listing~1 in the main paper) blocks downstream release when severity is medium or higher. The default iteration limit is \texttt{DEFAULT\_MAX\_REFINEMENT\_ITERATIONS = 2}, configurable at initialization.

\section{Phase 5: ABI Generator Agent}
\label{app:abi}

The ABI Agent (\textit{Ethereum ABI Specialist}) generates the complete JSON ABI array from refined Solidity code. Requirements enforced by the task prompt include:

\begin{itemize}
    \item Every \texttt{public}/\texttt{external} function must appear, including getter functions generated from public state variables.
    \item All parameters must use exact Solidity types (\texttt{uint256}, not \texttt{uint}; explicit array types; \texttt{uint8} for enums; struct fields expanded individually).
    \item Event parameters carry \texttt{"indexed": true} where declared indexed in Solidity.
    \item \texttt{stateMutability} is set correctly: \texttt{"pure"} (no state access), \texttt{"view"} (read only), \texttt{"payable"} (accepts ETH), or \texttt{"nonpayable"} (default).
    \item Parameter names are preserved exactly as written in the source.
    \item Output is the raw JSON array only; no markdown fences, no prose.
\end{itemize}

\section{Phase 6: MCP Server Agent}
\label{app:mcp}

The MCP Agent generates a complete Python FastMCP server exposing all contract endpoints as standardized AI-callable tools. The generated server uses Web3.py for blockchain interaction and loads ABI and environment variables from the same directory as the script. For each ABI function, a corresponding \texttt{@mcp.tool()} decorated function is generated: payable functions include the \texttt{value} field in the transaction, non-payable functions omit it, and view/pure functions use \texttt{contract.functions.X().call()} without building a transaction. All transaction tools sign with a private key from environment variables and return \texttt{\{"tx\_hash": hash\}}.

\section{Phase 7: Quality Evaluator Agent}
\label{app:evaluator}

The Quality Evaluator Agent (\textit{Smart Contract Quality Analyst}) performs five-dimensional scoring by reading the specification line by line, then inspecting the generated code for evidence. Scores are precise integers based on explicit point arithmetic, never rounded to the nearest 5.

\begin{table*}[t]
\caption{Smart Contract Evaluation Metrics}
\label{tab:evaluation_metrics}
\centering
\begin{tabular}{p{3.8cm} p{0.8cm} p{11cm}}
\toprule
\textbf{Metric} & \textbf{Wt.} & \textbf{Description} \\
\midrule
M1: Functional Completeness & 25\% & Exact and semantic function name matching (exact match +10 pts, semantic match +7 pts per function); implementation quality assessed on logic completeness (+5), access control (+3), event emissions (+2), and input validation (+2) per function. \\
\addlinespace[2pt]
M2: Variable/Parameter Fidelity & 15\% & State variable declaration, correct Solidity types, and active use in logic (60 pts); function parameter count, types, descriptive names, and active use (40 pts). \\
\addlinespace[2pt]
M3: State Machine Correctness & 15\% & Two scoring paths: Path~A (explicit FSM states in spec) evaluates state definition, transition implementation, and guard enforcement; Path~B (no explicit states) evaluates correctness of stateless design and access control quality. \\
\addlinespace[2pt]
M4: Business Logic Fidelity & 35\% & Obligations, financial logic, temporal constraints, conditional flows, and economic invariant enforcement. Most heavily weighted metric. \\
\addlinespace[2pt]
M5: Code Quality & 10\% & Absence of placeholders and TODOs; presence of NatSpec documentation; appropriate event emissions; overall code structure and organization. \\
\bottomrule
\end{tabular}
\end{table*}

\subsection{Metric Weights Rationale}
Business Logic (35\%) is most critical because incorrect economic or obligation logic defeats the contract's purpose. Functional Completeness (25\%) is heavily weighted since missing functions prevent required operations entirely. State Machine (15\%) and Variable Fidelity (15\%) are equally weighted, as incorrect state transitions produce invalid lifecycle behavior and variable fidelity ensures auditability and upgradability. Code Quality (10\%) covers important hygiene concerns that are secondary to semantic correctness.

\subsection{Scoring Formulas}
For Functional Completeness (M1), points are calculated as:
\begin{equation}
    M_1 = \frac{10 \times |ExactMatch| + 7 \times |SemanticMatch|}{|Expected| \times 10} \times 50 + Q_{impl}
\end{equation}
where $Q_{impl}$ is the implementation quality score (up to 50 points) based on logic completeness (+5), access control (+3), event emissions (+2), and input validation (+2) per function. Variable Fidelity (M2) allocates 60 points for state variable completeness and types and 40 points for function parameter accuracy. The composite score uses Equation~\ref{eq:composite} from the main paper and is always recomputed deterministically in post-processing.

\section{Finite State Machines in Smart Contracts}
\label{app:fsm}

Finite State Machines (FSMs) provide a formal foundation for specifying smart contract behavior. An FSM defines \textit{states} as discrete phases of a contract's lifecycle (e.g., \textsc{Active}, \textsc{Pending}, \textsc{Completed}), \textit{transitions} as conditions and actions triggering state changes, \textit{guards} as preconditions that must hold for a transition to fire, and \textit{actions} as on-chain operations executed during transitions.

In the FSM-SCG dataset, each entry includes a formal FSM specification alongside the natural language requirements and ground-truth Solidity implementation. The Requirement Specification Agent extracts this FSM structure into the \texttt{conditions} field of the \texttt{Universal\-Contract\-Schema} (state names, transitions, events, guard conditions), the Generator Agent enforces it through enum-based state variables and transition guards, and the Quality Evaluator's M3 metric directly scores the accuracy of the FSM implementation.

\section{System Components}
\label{app:components}

\subsection{Core Package Structure}
\begin{itemize}
    \item \texttt{agents.py}: All seven agent instantiations and the \texttt{should\_refine()} gate logic
    \item \texttt{translator.py}: \texttt{IBMAgenticContractTranslator} main orchestrator, responsible for streaming pipeline coordination across all phases
    \item \texttt{task\_builders.py}: Detailed task description constructors for each agent, including domain-specific contract type detection and mandatory requirement injection
    \item \texttt{programs.py}: Legacy IBM Agentics \texttt{Program} class wrappers for backward compatibility
    \item \texttt{schemas.py}: Pydantic data models (\texttt{Universal\-Contract\-Schema} and supporting classes)
    \item \texttt{solidity\_compiler.py}: Compilation checking via subprocess call to \texttt{solc}/\texttt{solcjs}, with pragma normalization to \texttt{>=0.4.0}
\end{itemize}

\vspace{120pt}
\subsection{Agent Configurations}
Each agent is instantiated with a \texttt{role}, \texttt{goal}, and
\texttt{backstory} that govern its behavior across all pipeline
invocations. Complete role and goal definitions for all seven agents
are documented in Appendix~\ref{app:prompts}. The structural pattern
is uniform across agents:

\begin{lstlisting}[language=Python,
  caption={Agent Instantiation Pattern},
  basicstyle=\ttfamily\scriptsize,
  breaklines=true, columns=flexible]
agent = Agent(
    role="<agent role title>",
    goal="<task objective and output format>",
    backstory="<domain expertise and behavioral constraints>",
    llm=crew_llm,
    verbose=False,
    allow_delegation=False
)
\end{lstlisting}

\section{Demo Application}
\label{app:demo}

The \texttt{launch\_demo.py} script provides an interactive demonstration environment. It starts two services: an HTTP server on port 8000 serving a \texttt{sampler.html} interface for browsing the dataset, and a Flask API on port 5000 handling live translation requests from a \texttt{demo.html} frontend. The system automatically opens both browser interfaces. Users can browse the dataset by contract type and complexity, select a specification, observe the complete pipeline execute with real-time streaming output, and see per-metric quality scores with identified strengths and weaknesses. This complements batch processing by providing detailed pipeline introspection for research and debugging.

\section{Configuration Options}
\label{app:config}

The \texttt{IBMAgenticContractTranslator} supports the following initialization parameters:
\begin{itemize}
    \item \texttt{model}: LLM selection (default: \texttt{gpt-4o-mini})
    \item \texttt{enable\_reinforcement}: Toggle the severity-gated reinforcement loop (default: \texttt{True})
    \item \texttt{enable\_deployment}: Deploy generated contracts to a Ganache testnet (default: \texttt{False})
    \item \texttt{max\_refinement\_iterations}: Maximum audit-refine cycles (default: \texttt{2})
\end{itemize}

\section{Agent System Prompts}
\label{app:prompts}

The following prompts define the role, goal, and behavioral backstory
for each agent. Goals and backstories govern how the underlying LLM
approaches each task; they are injected at agent-creation time and
persist across all invocations within a pipeline run.

\subsection{Parser Agent} \leavevmode

\begin{lstlisting}[breaklines=true, basicstyle=\scriptsize\ttfamily,
  caption={Parser Agent -- Role and Goal}]
role: "Contract Analysis Expert"

goal: "Extract every specific term, function name, variable name,
state name, party role, financial amount, and obligation from the
contract text exactly as written. Produce a fully-populated
UniversalContractSchema JSON object with no generic placeholders
and with obligations NEVER empty when functions or operations
are described."

backstory: "You are an expert contract analyst who reads every
sentence carefully. You extract EXACT terminology -- if the
contract says 'initializeLease' you write 'initializeLease',
not 'initialize'. You map every described operation to an
obligation with the correct authorized party. You never leave
the obligations array empty when functions are described."
\end{lstlisting}

\subsection{Generator Agent} \leavevmode

\begin{lstlisting}[breaklines=true, basicstyle=\scriptsize\ttfamily,
  caption={Generator Agent -- Role and Goal}]
role: "Senior Solidity Smart Contract Engineer"

goal: "Implement the EXACT contract specification provided in
the task. Read every MANDATORY requirement, every listed
obligation, and every domain-specific rule, then implement each
one completely with real on-chain logic. Produce a contract of
150-400 lines that fully and correctly satisfies the
specification."

backstory: "You are a senior Solidity engineer with deep
expertise in DeFi, tokens, governance, escrow, and marketplace
contracts. For every function you write, you ask: what
real-world operation does this represent, what invariant must
hold before and after, and what can go wrong? You enforce
economic invariants (token supply conservation, escrow balance
accounting), temporal logic (deadlines via block.timestamp),
and access control (every sensitive function has a
require()-backed modifier). You NEVER write empty functions,
placeholder comments, or stub implementations."
\end{lstlisting}

\subsection{Auditor Agent} \leavevmode

\begin{lstlisting}[breaklines=true, basicstyle=\scriptsize\ttfamily,
  caption={Auditor Agent -- Role and Goal}]
role: "Blockchain Security Auditor"

goal: "Identify every exploitable vulnerability in the Solidity
contract. For each issue, name the specific function affected
and describe the exact exploit path. Provide severity_level,
approved boolean, issues array, recommendations array with
line-level fixes, vulnerability_count, and security_score in
valid JSON."

backstory: "You are a blockchain security expert specializing
in Solidity smart contract audits. You methodically check for
reentrancy, access control gaps, integer overflow, timestamp
manipulation, locked ether, unbounded loops, and input
validation failures. Every issue you report names a specific
function and explains how an attacker could exploit it. Every
recommendation is a concrete code-level fix, not generic
advice. You return only valid JSON -- no markdown, no prose."
\end{lstlisting}

\subsection{Refiner Agent} \leavevmode

\begin{lstlisting}[breaklines=true, basicstyle=\scriptsize\ttfamily,
  caption={Refiner Agent -- Role and Goal}]
role: "Smart Contract Security Refiner"

goal: "Fix all identified security vulnerabilities in Solidity
smart contracts while maintaining original functionality."

backstory: "You are a Solidity security specialist who fixes
smart contract vulnerabilities. Given a contract and a list of
security issues from an audit, you rewrite the code to address
every vulnerability while maintaining the original
functionality. You follow the Checks-Effects-Interactions
pattern, add reentrancy guards where needed, implement proper
access control, validate all inputs with require(), and ensure
no silent failures. You return ONLY the fixed Solidity code."
\end{lstlisting}

\subsection{Quality Evaluator Agent} \leavevmode

\begin{lstlisting}[breaklines=true, basicstyle=\scriptsize\ttfamily,
  caption={Quality Evaluator Agent -- Role and Goal}]
role: "Smart Contract Quality Analyst"

goal: "Score the generated Solidity contract across five
metrics (functional completeness, variable fidelity, state
machine correctness, business logic fidelity, code quality).
Produce precise integer scores based on exact point
calculations -- never round to the nearest 5. Return only
valid JSON with metric_1 through metric_5 objects and a
composite_score."

backstory: "You are an expert smart contract quality analyst
who evaluates generated Solidity code against natural language
specifications. You read the specification line by line, then
inspect the code and assign scores based on exact evidence --
counting matched functions, checking that variables are written
and read, verifying state transitions are reachable, and
confirming economic invariants are enforced. Your scores are
precise (73 not 75) because you show the arithmetic. You
return only valid JSON -- no markdown, no prose."
\end{lstlisting}

\vspace{60pt}
\subsection{ABI Generator Agent} \leavevmode

\begin{lstlisting}[breaklines=true, basicstyle=\scriptsize\ttfamily,
  caption={ABI Generator Agent -- Role and Goal}]
role: "Ethereum ABI Specialist"

goal: "Generate the complete, accurate ABI JSON array for the
given Solidity contract. Include every public/external function
with correct inputs, outputs, and stateMutability; every event
with all parameters and indexed flags; and the constructor.
Types must be exact Solidity types (uint256 not uint). Return
ONLY the JSON array."

backstory: "You are an Ethereum developer who has spent years
generating and validating ABI specifications. You know that
'uint' must be 'uint256', that view functions have no state
mutations, and that indexed event parameters must carry
\"indexed\": true. You include every public/external function
-- never miss one. You return only the raw JSON array -- no
markdown fences, no prose."
\end{lstlisting}

\subsection{MCP Server Generator Agent} \leavevmode

\begin{lstlisting}[breaklines=true, basicstyle=\scriptsize\ttfamily,
  caption={MCP Server Generator Agent -- Role and Goal}]
role: "MCP Server Developer"

goal: "Generate production-ready MCP server code for
blockchain interaction."

backstory: "You are an expert Python developer specializing in
Web3.py and MCP server generation. You create complete,
self-contained FastMCP servers with proper error handling and
transaction management. For each ABI function you generate a
corresponding @mcp.tool() decorated function: payable
functions include value in the transaction, non-payable
functions omit it, and view/pure functions use .call()
without building a transaction. You load ABI and environment
variables from the script directory and return tx_hash for
all state-changing operations."
\end{lstlisting}

\begin{table*}[h!]
    \centering
    \small
    \begin{tabular}{|p{0.48\textwidth}|p{0.48\textwidth}|}
        \hline
        \multicolumn{2}{|c|}{\textbf{Example: Staking Contract Generation and Evaluation}} \\
        \hline
        \multicolumn{2}{|l|}{\textbf{Natural Language Specification (excerpt):}} \\
        \multicolumn{2}{|p{0.96\textwidth}|}{
            \textit{A staking contract where users can stake tokens during a farming period. The contract transitions through three states: Farming Not Started, Farming Ongoing, and Farming Ended. Users can stake tokens after the start time, withdraw staked tokens, and claim rewards based on their staking period. Rewards cease when the end time arrives.}
        } \\
        \hline
        \textbf{Generated Solidity Code (simplified)} & \textbf{Evaluation Results} \\
        \hline
        \begin{minipage}[t]{0.46\textwidth}
            \vspace{2pt}
            \begin{lstlisting}[basicstyle=\ttfamily\scriptsize, frame=none]
enum State {
  FarmingNotStarted,
  FarmingOngoing,
  FarmingEnded
}
State public currentState;
IERC20 public stakeToken;
mapping(address => UserInfo)
  public userInfos;
function stake(uint256 _amount)
  external afterStart beforeEnd {
  require(_amount > 0);
  stakeToken.transferFrom(
    msg.sender, address(this), _amount);
  userInfos[msg.sender]
    .amountStaked += _amount;
  emit Staked(msg.sender, _amount);
}
function withdraw(uint256 _amount)
  external nonReentrant {
  require(_amount > 0);
  userInfos[msg.sender]
    .amountStaked -= _amount;
  stakeToken.transfer(msg.sender, _amount);
  emit Withdrawn(msg.sender, _amount);
}
            \end{lstlisting}
            \vspace{2pt}
        \end{minipage}
        &
        \begin{minipage}[t]{0.46\textwidth}
            \vspace{2pt}
            \textbf{Composite Score:} 87.3/100 (Grade: B)

            \vspace{4pt}
            \textbf{Strengths:}
            \begin{itemize}[leftmargin=*, itemsep=0pt, parsep=2pt]
                \item All required functions (\texttt{stake}, \texttt{withdraw}, \texttt{claim}, \texttt{totalValue}) implemented with complete logic
                \item Explicit state machine with enum definition (\texttt{FarmingNotStarted}, \texttt{FarmingOngoing}, \texttt{FarmingEnded}) and transition guards
                \item Strong access control via time-bounded modifiers (\texttt{afterStart}, \texttt{beforeEnd}, \texttt{nonReentrant})
                \item Proper event emissions for all state changes
                \item Comprehensive input validation (\texttt{require(\_amount > 0)})
            \end{itemize}

            \vspace{4pt}
            \textbf{Weaknesses:}
            \begin{itemize}[leftmargin=*, itemsep=0pt, parsep=2pt]
                \item Missing \texttt{poolInfos} variable from specification
                \item No NatSpec documentation comments
                \item Compilation error due to interface placement
            \end{itemize}

            \vspace{4pt}
            \textbf{Metric Breakdown:}
            \begin{itemize}[leftmargin=*, itemsep=0pt, parsep=2pt]
                \item M1 Functional Completeness: 92/100
                \item M2 Variable Fidelity: 85/100
                \item M3 State Machine: 90/100
                \item M4 Business Logic: 86/100
                \item M5 Code Quality: 80/100
            \end{itemize}
            \vspace{2pt}
        \end{minipage} \\
        \hline
    \end{tabular}
    \caption{End-to-end example: staking contract generation from natural language. The pipeline extracts exact state names (\texttt{FarmingNotStarted}, \texttt{FarmingOngoing}, \texttt{FarmingEnded}), generates a complete Solidity implementation with proper state guards, and provides structured feedback identifying both successes and actionable deficiencies.}
    \label{tab:example_evaluation}
\end{table*}

\end{document}